\documentstyle[aps,prd,psfig]{revtex}
\begin{document}
\title{Electrons as Quasi-Bosons in Strong Magnetic Fields and
the Stability of Magnetars }
\author{Soma Mandal$^{a)}$ {\thanks{E-Mail: soma@klyuniv.ernet.in}}
 and Somenath Chakrabarty$^{a),b)}${\thanks{E-Mail: 
 somenath@klyuniv.ernet.in}} }
\address{
$^{a)}$Department of Physics, University of Kalyani, Kalyani 741 235,
India and
$^{b)}$Inter-University Centre for Astronomy and Astrophysics, Post Bag 4,
Ganeshkhind, Pune 411 007, India
}
\date{\today}
\maketitle

\begin{abstract}
Following the recent ideas of Dryzek, Kato, Mu\~noz and Singleton \cite{R1} 
(henceforth we call this paper as R1), that the composite system consisting 
of an electron along with its screened electric field and within the sphere 
of influence, the trapped  magnetic field of white dwarf can behave like a 
boson, we have argued that such an exotic transition (bosonization) of 
electronic component in strongly magnetized  neutron star matter in 
$\beta$-equilibrium can make the existence of magnetars physically viable.
\end{abstract}
One of the oldest subject of physics, "the study of the effect of strong 
magnetic field on dense matter of charged particles" has gotten a new 
dimension after the discovery of a few magnetars. These exotic objects are 
believed to be strongly magnetized young neutron stars of surface magnetic 
field $\approx 10^{15}$G and at the core region it may go up to $10^{18}$G 
\cite{R2,R3,R4,R5,R6}. It is therefore very much advisable to examine 
throughly the effect of such strong magnetic field on various physical 
properties of dense neutron star matter as well as on the physical processes 
taking place inside neutron stars.  An extensive studies have already been 
done on the equation of state of dense neutron star matter in presence of 
strong magnetic field.  Such studies are based on the quantum mechanical 
effect of strong magnetic field \cite{R7,R8}. The effect of strong quantizing 
magnetic field on the gross properties, e.g., the mass, radius, moment of 
inertia etc., of neutron stars, which are strongly dependent on the equation 
of state of matter have been obtained \cite{R9,R10}. How the weak 
processes, e.g., weak reactions and decays are affected by the quantum 
mechanical effect of strong magnetic field have also been derived from
the first principal \cite{R11}. 
These studies show that the $\beta$-equilibrium condition too depends on the 
strength of magnetic field. Since the cooling of neutron stars is dominated 
by the emission of neutrinos produced by weak processes inside the stars, 
these studies also give an idea of the effect of strong magnetic field on 
the thermal evolution of neutron stars. Not only that, the presence of strong 
magnetic field can change significantly, both qualitatively and quantitatively, 
the transport coefficients (e.g., viscosity coefficient, thermal conductivity, 
electrical conductivity etc.) of dense neutron star matter \cite{R11a,R12,R13}. 
The magnetic field can change the tensorial character of transport coefficients
of neutron star matter. Such qualitative changes in transport coefficients 
can cause significant changes in thermal evolution of neutron star 
matter and also the evolution of its magnetic field. There are another kind of 
studies; the effect of strong quantizing magnetic field on quark-hadron phase 
transition. It was shown explicitly that a first order quark-hadron phase 
transition is absolutely forbidden if the strength of magnetic field exceeds 
$10^{15}$G. However, a metal insulator type (color insulator to color metal) 
second order phase transition is possible unless the field strength exceeds 
$10^{20}$G \cite{R14,R15}. It has also been shown, that even if there is a 
first order quark-hadron phase transition for magnetic field strength  
$< 10^{15}$G at the core region of a neutron star, an investigation 
of chemical evolution of quark matter, with various initial conditions, 
leads to the system in $\beta$-equilibrium, revealed that the system becomes
energetically unstable in chemical equilibrium \cite{R16}. In a class of 
completely different type of studies, the mechanical stability and some of 
the gross properties of deformed stellar objects are analyzed with general 
relativity.  The presence of strong magnetic field can destroy the spherical 
symmetry of a neutron star \cite{R17,R18,R19}. Then it is quite possible for 
a deformed and rotating neutron star to emit gravity waves, which in principle 
can be detected \cite{R20}. It has also been shown with the help of general 
relativity that the presence of strong magnetic field may pose a serious 
problem on the stability of magnetars, in the extreme case it may become 
either a black disk or black strings \cite{R21}. In a very recent work, we 
have critically 
studied the ferro-magnetism of neutron star matter, which is believed to be 
one of the sources of residual magnetism of old neutron stars/sources  of 
magnetic field of millisecond pulsars. In these studies, we have shown 
explicitly that spontaneous ferromagnetic transition in absence of an external 
magnetic field is absolutely forbidden in neutron star matter in 
$\beta$-equilibrium \cite{R22}. However in the case of neutrino trapped 
neutron star matter (proto-neutron star matter), the possibility  of such a 
transition can not be ruled out, provided the neutrinos carry some non-zero 
mass.  We have also analyzed the problem of occupancy of only zeroth Landau 
levels by electrons/protons, which occur in presence of ultra-strong magnetic
fields \cite{R23}.  It has been argued in this microscopic model analysis 
that in presence of a strong quantizing magnetic field the existence of 
neutron star matter in $\beta$-equilibrium is questionable. Which further 
opens up a vital question on the possibility of magnetars as young and 
strongly magnetized neutron stars. The macroscopic general relativistic 
studies and the microscopic model calculations therefore arrive at the same 
conclusion- "the nonexistence of magnetars". To resolve this
controversy, we have
followed the ideas of reference  R1. We have also assumed that it is possible 
for the electrons to become quasi-bosons after combining with the screened 
electric field of the electron and the magnetic field of the star
trapped within the screened region. Since 
pions or kaons can not condense in presence of ultra strong magnetic field 
\cite{R24}, it is therefore not energetically favorable for the system to 
produce $\pi^-$ or $K^-$ instead of electrons. On the other hand, the 
condensation of quasi-bosonic states of composite electron system to almost 
zero kinetic energy state is energetically more favorable even with respect 
to the occupancy of only the zeroth Landau level for electrons. Now it is 
found in the literature that fermi-bose transition plays a significant role 
in various branches of physics, e.g.,  high energy particle physics,
and condensed matter physics. We therefore consider the
feasibility of bosonization of electrons inside the magnetars to solve
this serious  problem of instability of neutron star matter in 
$\beta$-equilibrium in presence of ultra strong magnetic field and hence
try to resolve the controversial issue of the existence of magnetars.

In the recent work (R1), the possibility of bosonization of electrons in 
magnetized white dwarfs has been proposed. In this paper, some of the 
interesting properties of white dwarfs relevant for such exotic transition 
have been discussed. The collapse of white dwarfs to neutron stars, because 
of bosonization has also been addressed in this reference.

Now several authors have investigated the fermi-bose transition during the last 
few decades in various branches of physics, e.g., in high energy particle 
physics, condensed matter physics - especially in strongly correlated electron 
system, fractional quantum Hall effect, anyon physics etc.

In the case of high energy particle physics, the application of fermi-bose 
transition has 
been studied in both abelian as well as in non-abelian field theories and 
also in non-linear Sine-Gordon model, in the low dimension and later, 
extended to $3+1$ dimensions \cite{R25,R26,R27}. In the condensed matter 
physics, there are certain systems in which fermions can be converted to 
quasi-bosons. The bosonization has been studied in detail for the crystalline 
as well as for the disordered condensed matter systems. Such studies are mainly 
concentrated in the low dimensional cases.  However, an extension to $3+1$ 
dimensions have also been done for certain cases. In $2D$, the anyons, which 
are the composite system of fermions and quantized magnetic flux, satisfy any 
statistics, have also been studied. The other interesting studies are on 
the strongly correlated electron system, the mesoscopic systems and the 
quantum Hall 
effect, and the high $T_c$ super-conductivity \cite{R28,R29,R30,R31,R32}.

The theoretical model presented in reference $(R1)$ is the first attempt to 
apply bosonization in the study of compact stellar objects. Following the 
ideas proposed in this reference, we shall try to convince that if electrons 
in magnetars are converted to quasi-bosons, the serious problem on the existence
of magnetars and also on the stability of dense neutron star matter in 
$\beta$-equilibrium in presence of extremely strong magnetic field 
posed by the electrons occupying only the zeroth Landau levels can in 
principle be solved. 
               
We assume that the electrons in magnetars can behave like quasi-bosons in 
combination with its screened coulomb field and the strong magnetic field 
trapped within the sphere of influence. The size of these composite objects 
is $\sim R_{sc}$ $\sim 10^{-11}-10^{-12}$ cm in the relativistic region, 
where $R_{sc}$ is the screening length of the electrostatic field of electrons.
To obtain the screening length we use the relativistic version of 
Thomas-Fermi model. According to which, the chemical potential of the 
electron is given by
	\begin{equation}
	\mu_e= (p_{F_e}^2(r)+m_e^2)^{1/2}-e\phi(r) ={\rm{constant}}
	\end{equation}
where $p_{F_e}(r)$, $m_e$ and $e$ are respectively the fermi momentum, mass and 
magnitude of charge of the electron and $\phi$ is the effective electrostatic 
field experienced by the electron.  The well known Poisson's equation is then 
given by 
	\begin{equation}
	\nabla^2 \phi= -4\pi en_p'(r)+4\pi en_e'(r)
	\end{equation}
where $n_p'$ and $n_e'$ are respectively the perturbed densities of proton 
and electron from their equilibrium values $n_p^0$ and $n_e^0$ 
satisfy the relation $n_p^0=n_e^0$, to fulfill the charge neutrality 
condition. Then it is quite obvious that in the perturbed case, the
total electron density is given by
	\begin{equation}
	n_e=n_e^0+n_e'
	\end{equation}
Now
	\begin{equation}
	n_e=\frac{1}{3 \pi^2}p_{F_e}^3=\frac{1}{3 \pi^2}
	\{(\mu_e+e\phi)^2-m_e^2\}^{3/2}
	\end{equation}
Hence we get for small $\phi$ approximation
	\begin{equation}
	n_e'(r)\approx \frac{3en_e^0\phi(r)\mu_e} {\mu_e^2-m_e^2}
	\end{equation}
On substituting $n_e'$ in the Poisson's equation, we have the screening length
	\begin{equation}
	R_{sc}=\left (\frac{\mu_e^{2}-m_e^{2}} {12\pi e^2n_e^0\mu_e}
	\right )^{1/2}
	\end{equation}
Now the angular momentum of electromagnetic field (combination of screened 
electrostatic field of electron and the strong magnetic field of the
magnetar trapped within this region) 
is given by (we have assumed $\hbar=c=1$)
	\begin{equation}
	\vec L^{em}= 2eBR_{sc}^2 \hat z
	\end{equation}
where $\hat z$ is an unit vector along Z-axis which is also assumed to be the 
direction of strong magnetic field $B$.	  Then the total spin of the 
composite system is given by
	\begin{equation}
	\vec S_{eff}= \vec S_e+ \vec L^{em}
	\end{equation}
If the eigen value of $L^{em}$ is an odd integer multiple of $1/2$, the 
composite system behaves like a quasi-boson.

To explain how the bosonization of electron gas depends on matter density 
and to 
obtain the corresponding critical value of magnetic field strength at which 
the composite system behaves like a quasi-boson, we put
	\begin{equation}
	L^{em}=\frac{s}{2}
	\end{equation}
where $s$ is a positive odd integer parameter. The quasi-bosons will therefore 
behave either like scalar, vector or tensor particles depending on the value 
of $s$.

In fig.1 we have plotted the critical strength of magnetic field against the 
baryon number density for various values of the parameter $s$. This figure 
shows that if there is possibility of any such composite objects of electrons 
along with its screened electro-static field and the magnetic field of the 
star trapped within the sphere of radius $R_{sc}$, then the quasi-bosons are
not necessarily of scalar or vector type, but they can have tensorial
character too of rank $\geq 2$, depending on the magnitude of
magnetic field strength. Although, theoretically predicted
upper limit for spin of boson is $2$, which is called graviton, it is in 
principle possible to have quasi-bosons (which are of course not real bosons) 
of spin $\geq 2$ inside the magnetars, especially, if the magnetic field at 
the core region $\sim 10^{18}$G, for which the maximum value of spin for 
quasi-boson states at the core region can be $ 4 \hbar$. Since these composite 
objects are not real bosons and also have finite dimension ($\sim R_{sc}$), 
then instead of comparing them with real bosons, it is better to draw some 
analogy with the high spin states of nuclei formed in nuclear fusion 
reactions. The finite dimension of these composite objects justifies 
to draw the comparison with high spin states of the nuclei. Whatever be the 
corresponding real world picture, these quasi-bosons should condense within the 
magnetar.  Since the stability of the neutron stars is governed by the 
degeneracy pressure of neutron matter, the condensation (transition to almost 
zero momentum state) of the quasi-bosons does not affect the mechanical 
equilibrium of the system.  Further, the problem arises out of the occupancy 
of only the zeroth Landau level by electrons is also solved. As a consequence, 
the stability of strongly magnetized dense neutron star matter in 
$\beta$-equilibrium is gurrented.  Now there are two different kinds of
general relativistic studies on the stability of strongly magnetized
neutron stars. In purely classical studies, it has been shown that the
inward magnetic pressure along the magnetic axis deforms the
spherically symmetric structure of a star to an oblate shape. In the
extreme case, it can become a black disk. Which of course needs a very
high magnetic field ($\geq 10^{20}$G). If it is so, the instability of
strongly magnetized neutron stars as predicted by classical general
relativity can not be solved with microscopic (bosonization) model
described in this article. In the other type of general relativistic
studies, the effect of paramagnetism of Landau diamagnetic system has
been considered. Now the paramagnetic behavior of Landau diamagnetic
system dominates if only the zeroth Landau levels are occupied by the
charged fermions. In this case all the spins are aligned. In such a
scenario, the spherically symmetric structure of the star will be
distorted to prolate shape and in the extreme case, it takes the shape
of a cigar or reduces to a black string (one dimensional object). Since
this kind of instability is connected to the paramagnetic behavior of
charged fermions (mainly electrons) occupying Landau levels, the
possibility  of bosonization of electronic components can solve this
problem. Which therefore  removes the controversy on the non-existence
of magnetars.

\begin{figure} 
\psfig{figure=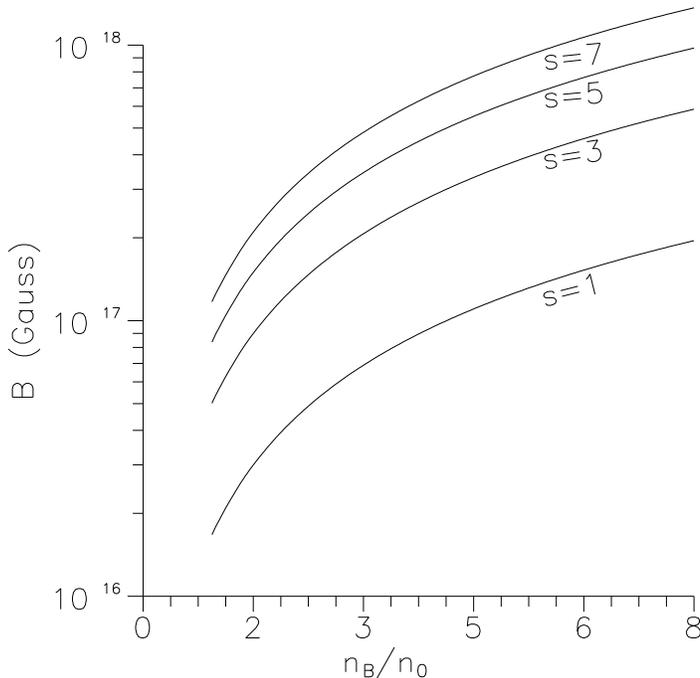,height=0.5\linewidth}
\caption{
Variation of magnetic field with the baryon number density (expressed in
terms of normal nuclear density) for various values of the parameter
$s$.
}
\end{figure}

\noindent {\sl{Acknowledgment: SC is thankful to the Department of Science and 
Technology, Govt. of India, for partial support of this work, Sanction 
number:SP/S2/K3/97(PRU).}}
\end{document}